\documentstyle[twocolumn,prl,aps,epsfig]{revtex}
\newcommand{\capt}[1]{\caption[*]{\small #1}}

\begin{document}
\draft
\twocolumn[\hsize\textwidth\columnwidth\hsize\csname @twocolumnfalse\endcsname
%
%
%

\title{Thermodynamics of the one-dimensional SU(4) symmetric spin-orbital 
model}

\author{Beat Frischmuth $^1$, Fr\'ed\'eric Mila $^2$, Matthias Troyer $^3$}
\address{$^1$ Institute of Theoretical Physics, ETH H\"onggerberg, 
CH-8093 Z\"urich, Switzerland \\
$^2$ Laboratoire de Physique Quantique, Universit\'e Paul Sabatier, 118 
route de Narbonne, F-31062 Toulouse Cedex \\
$^3$ Institute for Solid State Physics, University of Tokyo, Roppongi 7-22-1,
Tokyo 106 }

\date{\today}
\maketitle

\begin{abstract}
  The ground state properties and the thermodynamics of the
  one-dimensional SU(4) symmetric spin system with orbital degeneracy
  are investigated using the quantum Monte Carlo loop algorithm.  The
  spin-spin correlation functions exhibit a 4-site periodicity, and
  their low temperature behavior is controlled by two correlation
  lengths that diverge like the inverse temperature, while the entropy
  is linear in temperature and its slope is consistent with three
  gapless modes of velocity $\pi/2$. The physical implications of
  these results are discussed.
\end{abstract}

\vskip2pc]
\narrowtext

In many transition metal oxides, the electron configuration has an
orbital degeneracy in addition to the spin degeneracy. The sign and
magnitude of the spin-spin interactions is then determined by the
orbital occupation leading to strong coupling between orbital and spin
structure (for an overview see Ref. \cite{riceOrbital}). The Hamiltonian
describing such spin-1/2 systems with two-fold orbital degeneracy
(isospin $\tau=1/2$) was derived by Castellani and coworkers more than
20 years ago \cite{castellani}. The Hamiltonian has rotation symmetry in
$\vec{S}$-space. In $\vec{\tau}$-space this symmetry is broken by a
Hund's rule term. Recently, the investigation of these spin-orbital
models has attracted renewed interest, following the progress in the
experimental studies of transition metal oxides \cite{bao}.

In this Letter we study the Hamiltonian derived by Castellani {\em et
  al.} on a 1D chain, but neglecting the Hund's rule term. In this
isotropic case the Hamiltonian is:
\begin{equation}\label{ham1}
H=J\sum_i \left(2\vec{S}_i\cdot\vec{S}_{i+1}+\frac{1}{2}\right)
          \left(2\vec{\tau}_i\cdot\vec{\tau}_{i+1}+\frac{1}{2}\right).
\end{equation}
It is rotationally invariant not only in $\vec{S}$-space, but also in
$\vec{\tau}$-space. Furthermore it has an interchange symmetry between
spins and orbitals. In such a case, the standard mean-field
approach\cite{castellani} that leads to ferromagnetic correlations for
one type of variables and antiferromagnetic correlations for the other
one, should not be appropriate. Our main motivation is to study the
consequences of this symmetry, in more detail.

A number of analytic results have already been obtained on this model.
The system considered here (Eq. (\ref{ham1})), belongs to a class of
models which is exactly solvable in one dimension by the Bethe Ansatz.
The Bethe Ansatz solution obtained by Sutherland gives the exact ground state
energy and the ``spin wave'' excitations as well \cite{sutherland}.
For the model of Eq.  (\ref{ham1}), there are 3 gapless modes, having
all a common velocity $v=\pi J/2 $.  They are shown in Fig.~3 of
Ref.~\cite{sutherland}.

Second, it was pointed out very recently\cite{zhang,japan}, that the
Hamiltonian $H$ has not only the obvious SU(2)$\times$SU(2) symmetry,
but that the full symmetry of Eq. (\ref{ham1}) is the even higher
symmetry group SU(4). SU($N$) symmetric models in one dimension
were studied by Affleck, using conformal field theory \cite{affleck}.
He showed that any one-dimensional system of SU($N$) symmetry is
critical. He calculated explicitly the critical exponents and zero
temperature correlations and showed that at the very low energy scale
these models are equivalent to $N-1$ free massless bosons. These
general results naturally also applies to our case with $N=4$.

In this Letter we present the first investigation of the thermodynamic
properties of the model Eq. (\ref{ham1}). For this purpose we have
adapted the continuous time quantum Monte Carlo (QMC) loop algorithm
\cite{evertz} to spin-orbital models. In fact, this algorithm is so
powerful that we can also use it to investigate the zero-temperature
properties of the model: Systems of length $L=100$ (with periodic
boundary conditions (PBC)) and inverse temperatures $\beta J=400$ or
$800$ ($\beta\gg L/v$) are predominantly in the ground state, and the
small contributions from thermally excited states negligible compared
to our statistical errors. In this way, the ground state properties
can be investigated.  We start with a brief summary of these results
since some of them differ significantly from the density matrix
renormalization group (DMRG) results reported in Ref.\cite{japan}.

\begin{figure}[t]
  \begin{center}
  \epsfxsize=90mm
  \hspace*{-5mm}
  \epsffile{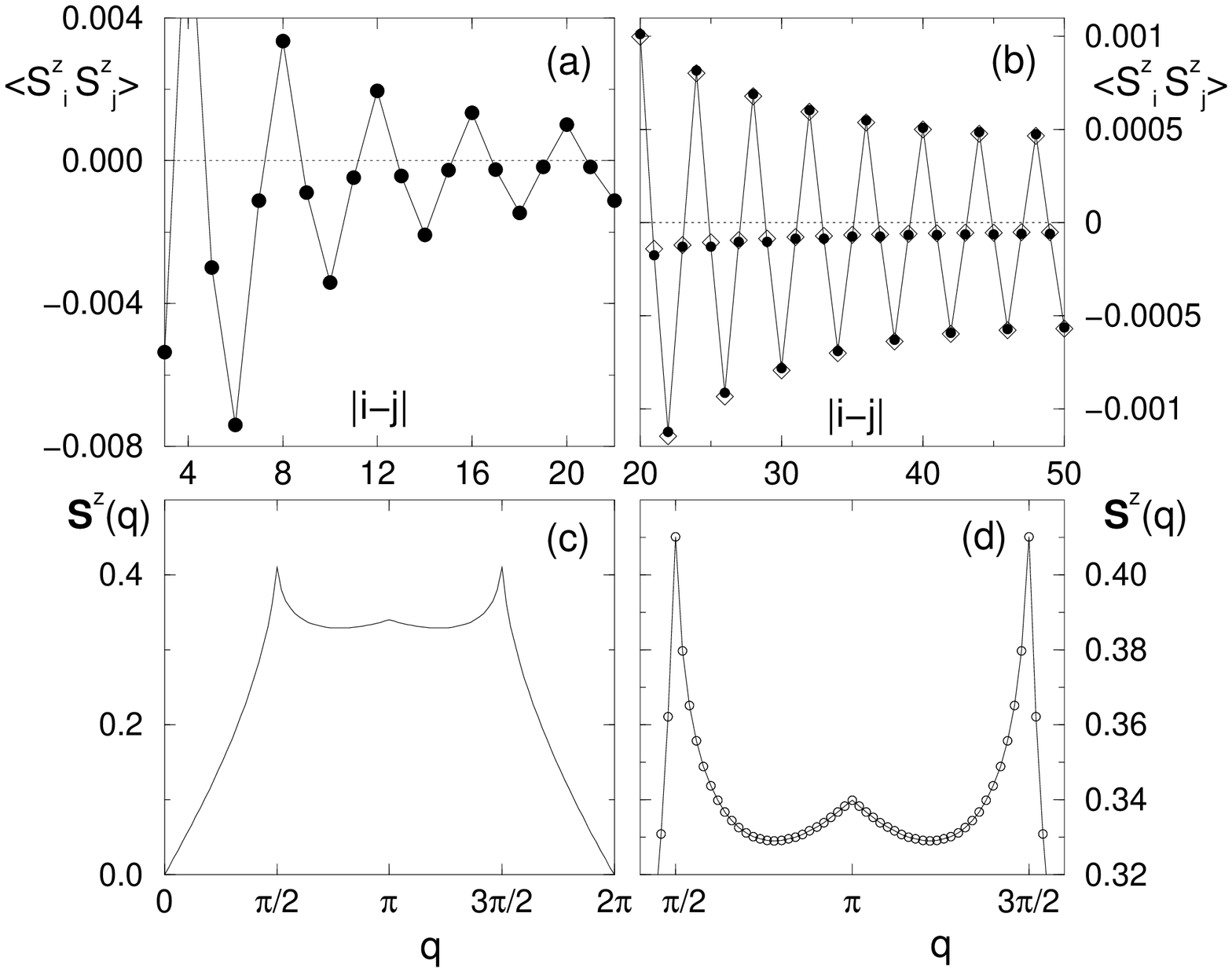}
  \end{center}
  \vspace*{-5mm} \capt{(a) QMC results for the correlation function
    $w_{ij}\equiv\langle S_i^z S_j^z\rangle$ (solid points) as a
    function of $|i-j|$ for a chain of length $L=100$ with PBC which
    is predominantly in the ground state (for details see text). The
    correlations for $|i-j|=1,\,2$ and 4 (which are out of the plot
    range) are -0.07168(1), -0.04011(1) and 0.008261(4), respectively.
    Fig.~(b) shows the correlations $w_{ij}$ at large distances
    $|i-j|$ and the fit (for details see text) to the QMC data
    (open diamonds). The statistical error bars of the QMC
    calculations are much smaller than the symbols. (c) and (d) show
    the Fourier transform ${\cal S}^z(k)$ of $w_{ij}$ on two different
    scales. \label{corr0RS}}
\end{figure}

The ground state energy for a chain of $L=100$ with periodic boundary
condition is found to be $\epsilon_0(L=100)=0.8253(1)$, in perfect
agreement with the Bethe Ansatz result for the infinite chain
($-0.8251189\ldots$) \cite{sutherland}.  The zero-temperature
correlation functions $w_{ij}(T=0)\equiv\langle S_i^z
S_j^z\rangle(T=0)$ as a function of $|i-j|$ (for $L=100$) are shown in
Fig.~\ref{corr0RS}ab and its Fourier transform in
Fig.~\ref{corr0RS}cd.

Note that according to the SU(4) symmetry, all the following
correlations are equal \cite{zhang}:
\begin{equation}\label{symmetry}
\langle S_i^\alpha S_j^\alpha \rangle = 
\langle \tau_i^\alpha \tau_j^\alpha \rangle =
\langle 4 S_i^\alpha S_j^\alpha \tau_i^\beta \tau_j^\beta \rangle=
w_{ij},
\end{equation}
independent of the indices $\alpha,\,\beta=x,y,z$. This relation is
valid for zero as well as for finite temperatures. While the first
equality also holds for an arbitrary SU(2)$\times$SU(2) symmetric
model with exchange symmetry of the $\vec{S}$ and
$\vec{\tau}$-variable, the second one is a special property of the
SU(4) symmetric model. All the QMC results have been checked for the
symmetry relation Eq. (\ref{symmetry}) and perfect agreement within
the statistical error has been found.

\vspace*{10mm}

The correlation function $w_{ij}$ shows a clear four-site periodicity
(see Fig.~\ref{corr0RS}). Its sign is positive if $|i-j|=4N$, $N$
integer and negative otherwise. The reason for the latter is the
tendency for every four neighboring sites to form a SU(4) singlet
\cite{zhang}.  Furthermore, from Fig.~\ref{corr0RS}, it can be seen
that the correlations for distances $|i-j|=4N$ and $4N+2$ decay much
slower than for $|i-j|=4N+1$ and $4N+3$. The explanation of this fact
is simple: The system considered here has low lying excitations at
$k=0,\,\pi/2$ and $\pi$ (see Fig.~3 of \cite{sutherland}) each of them
leading to a mode with wave vector $k$ in the long distance
correlations. The amplitudes of this modes are expected all to decay
according to a power law, but with different critical exponents
$\alpha_k$. From the results for $w_{ij}$ (Fig.~\ref{corr0RS}), it
can be concluded that the two dominant modes are those with $k=\pi/2$
(positive prefactor) and $k=0$ (negative prefactor). This is also
reflected in the Fourier transform ${\cal S}^z(k)$ of the correlation
function $w_{ij}$, having a characteristic cusp structure at
$k=0,\,\pi/2$ and $\pi$ (see Fig.~\ref{corr0RS}cd). While the cusps at
$k=0$ and $\pi/2$ are quite sharp, the one at $k=\pi$, however, is not
so pronounced, indicating that the $k=\pi$ mode is of all the three
the least dominant mode in the correlation function.

The two critical exponents $\alpha_{\pi/2}$ and $\alpha_0$ can be
determined from the QMC data of the real space correlation
function $w(r)\equiv w_{ij,\,|i-j|=r}$. Fitting $w(r)$ to the form 
$b_{\pi/2}(r^{-\alpha_{\pi/2}}+(L-r)^{-\alpha_{\pi/2}})
\cos(\frac{\pi}{2} r) + b_0 (r^{-\alpha_0} +
  (L-r)^{-\alpha_0})$ for the range $20\lesssim r\lesssim 50$
(making explicit use that our system has PBC), we find 
\begin{equation}\label{expQMC}
\alpha_{\pi/2}=1.50\pm0.01,\ \ \ \alpha_0= 1.85\pm0.16.
\end{equation}
The best fit is obtained for $b_{\pi/2}=0.091,\ \alpha_{\pi/2}=1.499,\ 
b_0=-0.035,\ \alpha_0=1.85$ and is shown in Fig.~\ref{corr0RS}.  A
precise estimate of $\alpha_0$ is not simple since the $k=0$ mode is
only a relative small superposition on the top of the much stronger
$k=\pi/2$ mode. The exponent $\alpha_{\pi/2}$, however, can be
determined to high precision.  These results are in very good
agreement with the prediction of Affleck, who calculated the critical
behavior of the SU(4) correlation function in an arbitrary SU(4)
symmetric model using conformal field theory \cite{affleck}. This
correlation function is proportional to $w_{ij}$, as a consequence of
the symmetry relation Eq. (\ref{symmetry}) and the exact results are
$\alpha_{\pi/2}=\frac{3}{2}$ and $\alpha_0=2$. The exponent
$\alpha_{\pi/2}$ has also been estimated, using DMRG
($\alpha_{\pi/2}\simeq 1.5\sim 2$) \cite{japan}. The DMRG results are
in principle more precise than the QMC results, but finite size
effects in DMRG studies are much bigger due to the use of open
boundary conditions.  Thus it is not surprising that our estimate
Eq.~(\ref{expQMC}) is much more precise.

At finite temperatures, the dominant components in the correlation
function, $w_{ij}(T)\equiv\langle S_i^z S_j^z\rangle (T)$ (note that
Eq. (\ref{symmetry}) holds also at finite $T$) which result from the
soft modes at $k=0$ and $\pi/2$, no longer decay according to a power
law, but exponentially. The corresponding correlation lengths $\xi_0(T)$
and $\xi_{\pi/2}(T)$ may be different. 

\begin{figure}[h]
  \begin{center}
  \epsfxsize=85mm
  \epsffile{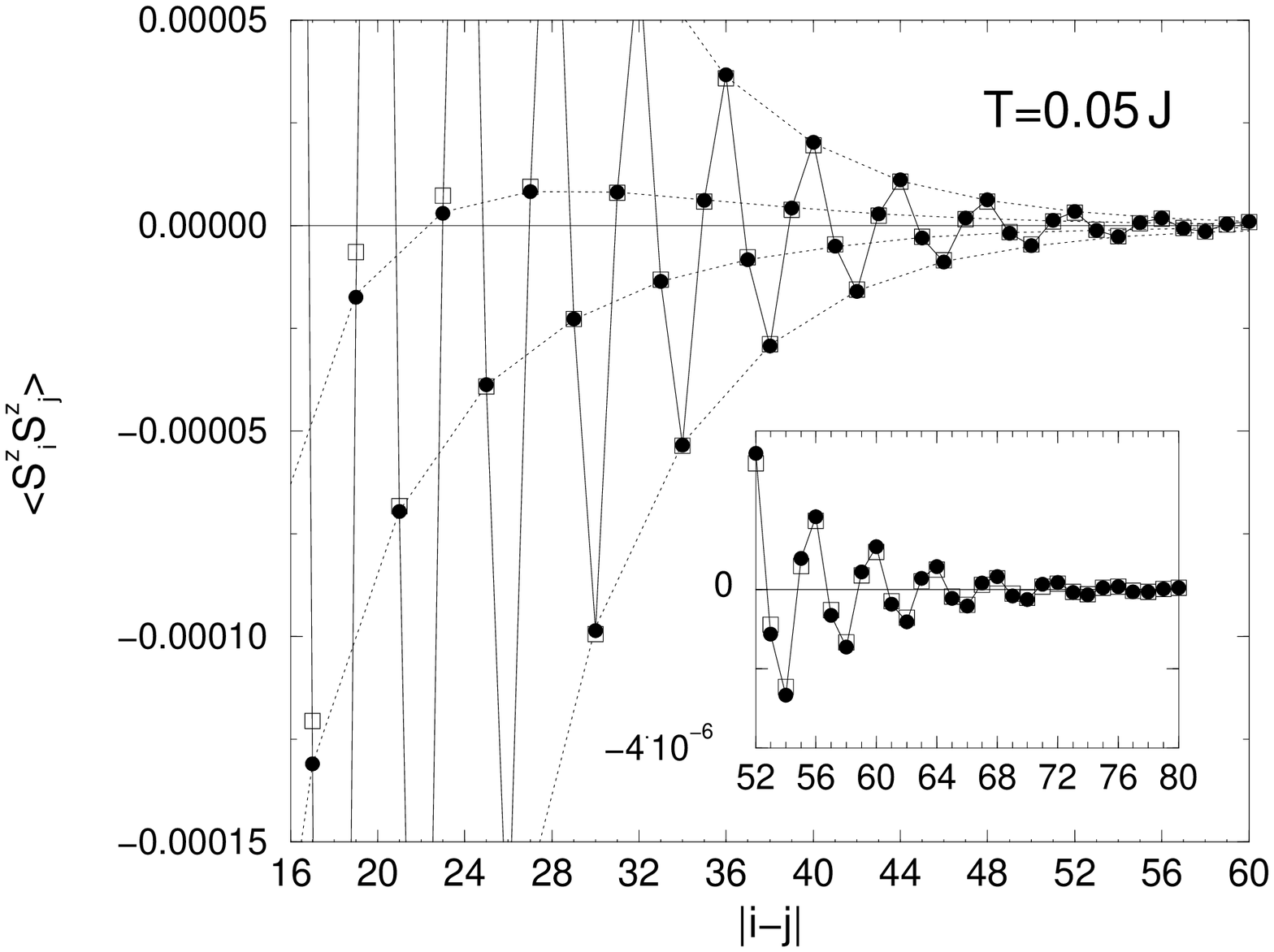}
  \end{center}
  \capt{QMC results for $w_{ij}(T)\equiv\langle
    S_i^zS_{j}^z\rangle(T)$ for a system of length $L=200$ with PBC
    and at temperature $T=0.05J$ (solid points). As a guide of the
    eye, the correlations at distances $|i-j|=4N+m$ are connected by a
    separate (dotted) line for each $m=0,1,2$ and 3.  The inset shows
    the long distance correlations. The error bars are much smaller
    than the symbols. The open squares show the fit of form Eq.
    (\ref{corrTform}) to the QMC data. 
    \label{corrTRS}}
\end{figure}

The correlation function $\langle S_i^z S_j^z\rangle(T)$ is shown as a
function of $|i-j|$ in Fig.~\ref{corrTRS} for a system of length
$L=200$ with PBC at a temperature $T=0.05J$. To find the correct
low-temperature form, describing the long distance behavior ($|i-j|\gg
\xi_0,\xi_{\pi/2}$) of the correlations $w_{ij}(T)$, one has to
consider not only a phase shift $\delta(T)$ in the $k=\pi/2$ mode, but
also an incommensuration effect of this component, i.e. that the
period is shifted away from $k=\pi/2$ by an amount $\phi_k(T)$. This
is due to the asymmetry of the excitation spectrum at the point
$k=\pi/2$ which manifests itself at finite $T$, where also excited
states contribute to $w_{ij}(T)$.  This asymmetry can be seen in
Fig.~2 of \cite{japan}, where the degeneracy of the lowest ``spin
wave'' branch is indicated.  As the degeneracy for $k>\pi/2$ is larger
than for $k<\pi/2$, we expect the weight of the $\pi/2$ mode to be
shifted to a higher $k$ value.  This effect can also be observed in
the Fourier transform ${\cal S}^z(k,T)$ of the correlation function,
where the maximum at $k=\pi/2$ at $T=0$ moves to higher $k$-values
when $T$ increases.

Finally, we propose the following low temperature form for the
correlations $w_{ij}(T)$ with $|i-j|\gg \xi_0,\xi_{\pi/2}$:
\begin{eqnarray}\label{corrTform}
w_{ij}(T)=b_0(T) e^{-|i-j|/\xi_0(T)}
+ b_{\pi/2}(T) e^{-|i-j|/\xi_{\pi/2}(T)} \nonumber\\
\quad\cos[(\pi/2+\phi_k(T)) (i-j)+\delta(T)].
\end{eqnarray}

Fitting the above form to the QMC data of $w_{ij}(T)$ for various $T$
gives the temperature dependence of the six parameters
$b_0,\,\xi_0,\,b_{\pi/2},\,\xi_{\pi/2},\,\phi_k$ and $\delta$. The
corresponding fit for $T=0.05J$ is shown in Fig.~\ref{corrTRS}. At
this point, we want to emphasize that it is important to include the
effect of incommensuration (i.e. including a parameter $\phi_k\neq
0$) to get accurate fits in the temperature range $0.01J \leq T \leq
0.08J$ and that at finite temperatures and larger distances, also the
correlations at distances $4N+3$ can become positive
(Fig.~\ref{corrTRS}), different from the correlations at $T=0$.  For
all considered $T$, the correlation lengths are much smaller
than the length of the system ($L>6\xi_{\pi/2},\,6 \xi_0$), so that
finite size effects are negligible.

\begin{figure}[h]
  \begin{center}
  \epsfxsize=85mm
  \epsffile{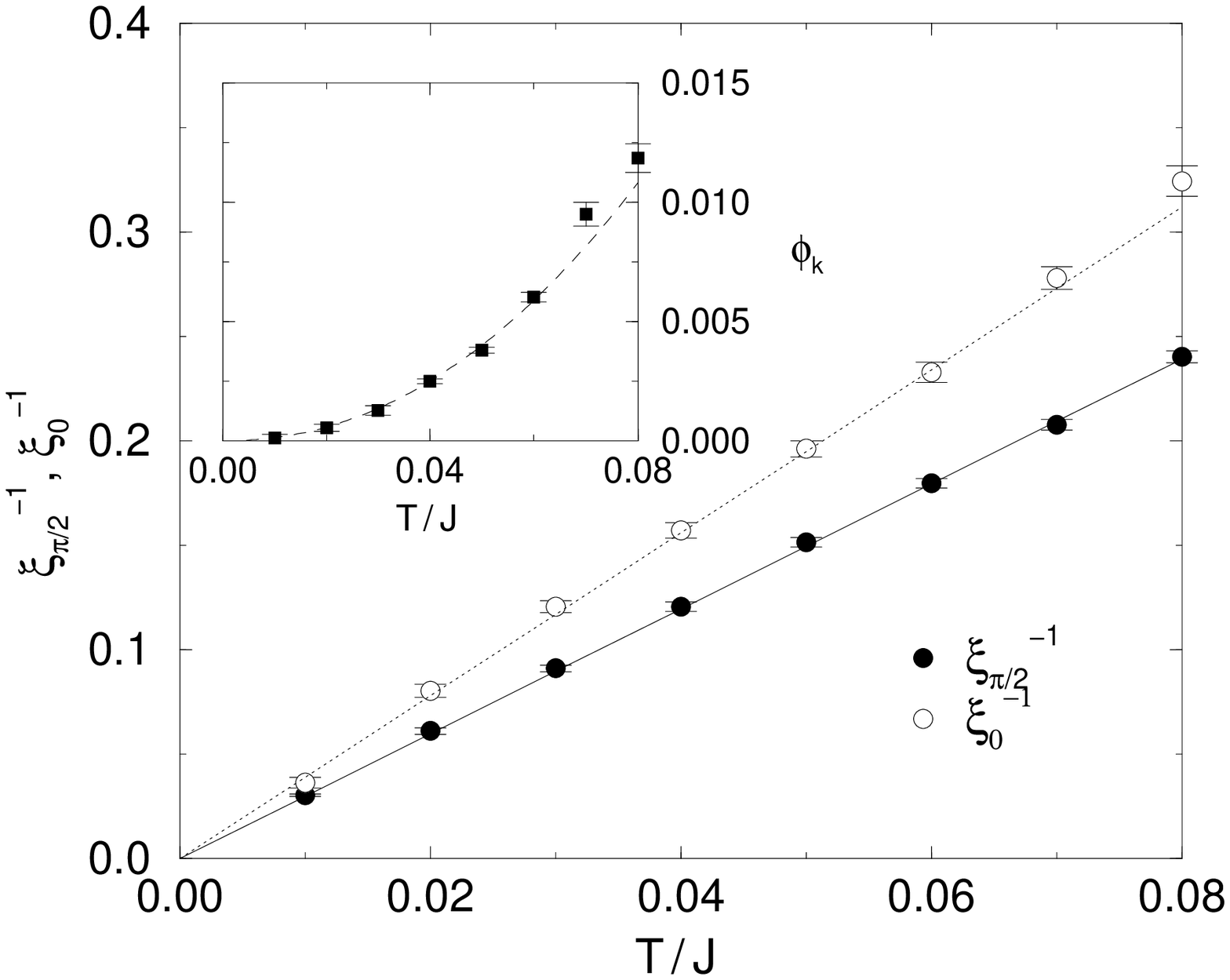}
  \end{center}
  \capt{Inverse correlation lengths $\xi_0^{-1}$ (open circles) and
    $\xi_{\pi/2}^{-1}$ (filled circles) and the periodicity shift
    $\phi_k$ (inset) as a function of temperature.  The leading $T$
    dependences (Eq. (\ref{leadT})) are also shown (solid, dotted and
    dashed lines).\label{tempdep}}
\end{figure}

In Fig.~\ref{tempdep} the inverse correlation lengths $\xi_0^{-1}$ and
$\xi_{\pi/2}^{-1}$ as well as the periodicity shift $\phi_k$ are
plotted as a function of temperature. At very low temperatures,
$\xi_0^{-1}$ and $\xi_{\pi/2}^{-1}$ both show a linear behavior (see
Fig.~\ref{tempdep}) and the leading temperature dependence are found
to be
\begin{equation}\label{leadT}
\xi_{\pi/2}^{-1}= (2.99 \pm 0.03)T,\,\,\,\,\,\,\,
\xi_0^{-1}= (3.90 \pm 0.09)T 
\end{equation}
Therefore both correlation lengths scale with $1/T$. This scaling
behavior, including the prefactor, can be motivated in the following
way.  By the Lorentz invariance of the underlying field theory of
the considered model and by exchange of the imaginary time and space
direction, one has $\xi_k(T)=v/\Delta_k(L=v/T)$, where $\Delta_k(L)$
is the finite size gap to the lowest excitations at wave vectors
$\approx k$ in a system of length $L$. $v$ is the spinon velocity,
which in our model is $\pi J/2$. For $k\approx 0$, the lowest lying
excitation energy is $\Delta_0(L)=v\cdot2\pi/L$ leading to
$\xi_0=J/(4T)$, in good agreement with Eq.~(\ref{leadT})).  For
$k=\pi/2$, the finite size results of Ref.~\cite{japan} show that
$\Delta_{\pi/2}(L)\approx 0.75 \Delta_{0}(L)$ (a similar result should
be obtained, using field theory) and hence $\xi_{\pi/2}\approx
J/(3T)$, again in very good agreement with Eq.~(\ref{leadT}). In
comparison, the leading temperature dependence of the correlation
length in the SU(2) AF Heisenberg model ($H_{\rm HB} =J \sum_i
\vec{S}_i\cdot\vec{S}_{i+1}$, $v_{\rm HB}=\pi J/2$ and $\Delta^{\rm
  HB}_\pi(L)=\pi v/L$) is $\xi_{\pi}^{\rm HB}(T)=J/(2T)$.

The leading $T$ dependence of the periodicity shift $\phi k$ is
determined by fitting $\lambda T^\beta$ to the data at very low
temperature $\phi_k(T) \propto T^{\,2.11\pm0.15}$. This scaling
exponent is quite close to the value of 2, which one would expect from
a simple calculation, considering the thermal admixtures of the spin
wave branches.

\begin{figure}[h]
  \begin{center}
  \epsfxsize=85mm
  \epsffile{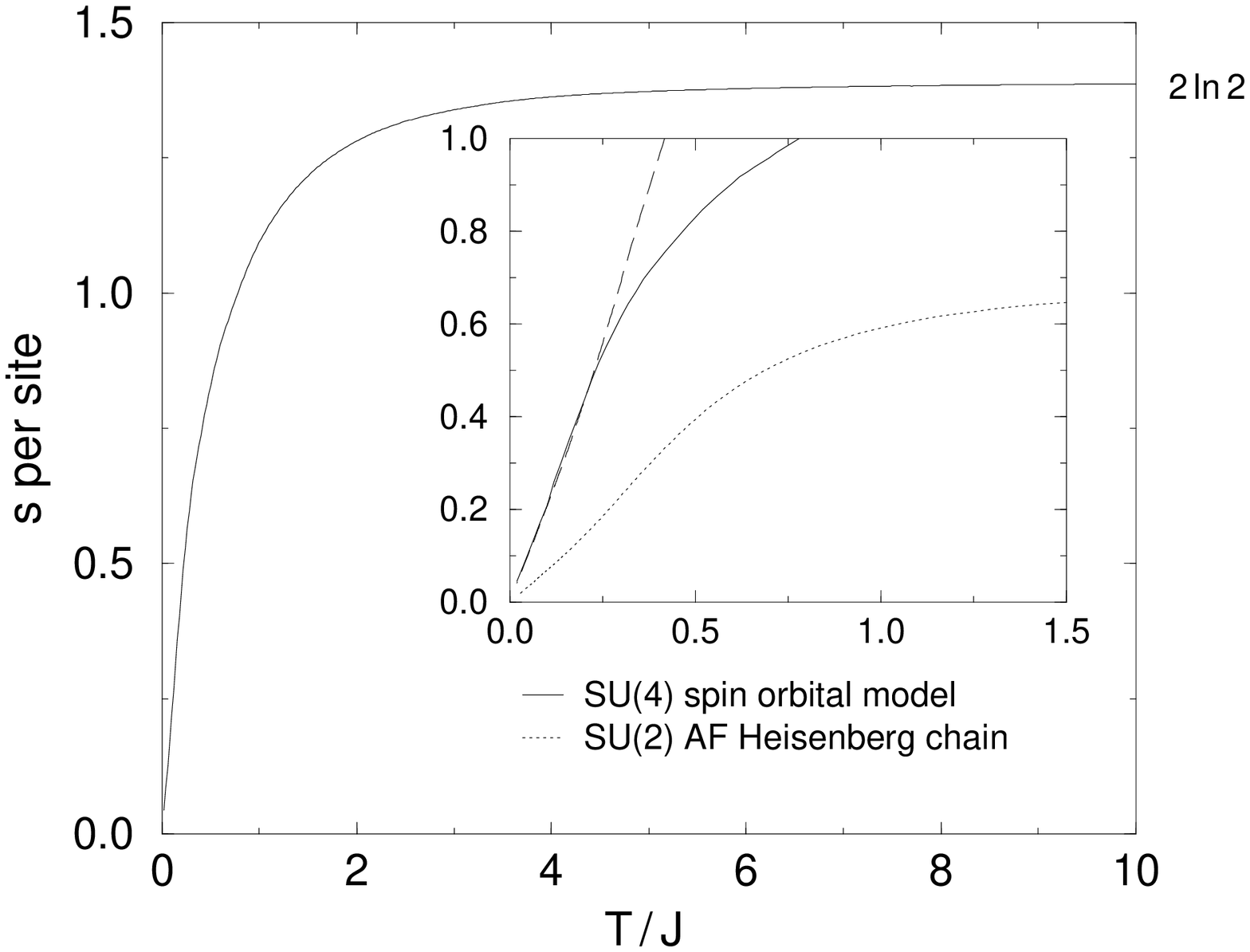}
  \end{center}
  \capt{Temperature dependence of the entropy $s$ per site for the
    spin-orbital model Eq. (\ref{ham1}) (solid line). In the inset the
    entropy per site is shown on larger temperature scale together
    with the entropy $s_{\rm HB}$ per site of a SU(2) spin-1/2 AF
    Heisenberg chain ($H_{\rm HB}$) (dotted line). For comparison also 
    $3s_{\rm HB}$ is shown (dashed line).\label{entropy}}
\end{figure}

Finally, we concentrate on the entropy $s$ per site of the SU(4)
invariant model Eq. (\ref{ham1}). Its $T$-dependence $s(T)$ is shown
in Fig.~\ref{entropy}. With decreasing $T$, the entropy decreases
monotonically from the high temperature value $\ln 4$ to 0 at zero
temperature. At low temperatures the entropy shows a linear behavior
as in the AF SU(2) Heisenberg chain ($H_{\rm HB}$). The
slope in the spin orbital model, however, is about a factor {\em
  three} bigger than that in the AF Heisenberg chain (see inset of
Fig.~\ref{entropy}). This is consistent with the statement of Affleck
\cite{affleck} that the AF Heisenberg model is equivalent to {\em one}
free massless boson, while the SU(4) invariant spin orbital model is
equivalent to {\em three} massless bosons. The velocity of these
bosons are all equal to $\pi J/2 $ \cite{sutherland,cloiseaux}.
Therefore we expect the low energy density of states (and hence the
entropy) of these two models just to differ by a factor 3.

The implications of these results for mean field treatments are far
reaching. To put them in perspective, it is useful to compare them to
the standard mean-field decoupling\cite{castellani}
$(\vec{S}_i\cdot\vec{S}_{i+1})(\vec{\tau}_i\cdot\vec{\tau}_{i+1})$
$\rightarrow$ $<\vec{S}_i\cdot\vec{S}_{i+1}>
\vec{\tau}_i\cdot\vec{\tau}_{i+1}$
$+<\vec{\tau}_i\cdot\vec{\tau}_{i+1}>\vec{S}_i\cdot\vec{S}_{i+1}$
$-<\vec{S}_i\cdot\vec{S}_{i+1}><\vec{\tau}_i\cdot\vec{\tau}_{i+1}>$.
Such a decoupling has a number of consequences. First of all, the
correlation function $\langle (\vec{S}_i\cdot\vec{S}_{i+1})
(\vec{\tau}_i\cdot\vec{\tau}_{i+1})\rangle$ should be equal to the
product of $<\vec{S}_i\cdot\vec{S}_{i+1}>$ with
$<\vec{\tau}_i\cdot\vec{\tau}_{i+1}>$, in clear contradiction both
with the fact that all of them are negative according to our results
and with the property of Eq. (3). Besides, and more importantly, if
such a decoupling was a valid approximation, the low-lying excitations
should consist of two branches corresponding to spin and orbital
excitations, respectively. This is again in clear contradiction with
the 3 low-lying modes of the Bethe ansatz which control the low
temperature physics according to our entropy results. So there is a
manifest break-down of the mean-field decoupling when spin and orbital
degrees of freedom play a symmetric role.

What is then the nature of the low-lying excitations? A full answer
cannot be given on the basis of the present results, but a number of
conclusions can be reached. Let us start with the 2-site problem.  The
ground state is six-fold degenerate
(spin-triplet$\times$orbital-singlet or
spin-singlet$\times$orbital-triplet), and energy may
be gained by allowing fluctuations between these local configurations.
The mean-field decoupling fails because it cannot take advantage of
these fluctuations.  Elementary considerations show that the best
mean-field decoupling leads to a very poor
estimate of the ground state energy (-0.3863 vs -0.8251 for the exact
result). That it is possible to gain energy by allowing the system to
fluctuate locally is best exemplified by the 4-site problem.  In fact,
the exact ground state for a 4-site cluster with periodic or open
boundary conditions, the SU(4) singlet of Ref.\cite{zhang}, can be
obtained exactly in terms of these dimer wave functions, and the
energy per bond is equal to -1, i.e. each bond has now managed to
reach its ground state energy thanks to the fluctuations between these
6 local configurations. Note that this is no longer true for longer
systems, indicating that 4-site clusters should be a good starting
point for building variational wave-functions. This can be seen as
the physical origin of the 4-site periodicity of the correlation
functions. A similar conclusion was reached in Ref.\cite{zhang} on the
basis of the SU(4) symmetry.

Finally, if the ground state is an RVB-like state involving resonances
between different local configurations, we are lead to the conclusion
that the elementary excitations cannot be pure spin or orbital
excitations, but composite objects where spin and orbital degrees of
freedom are intimately connected. Work is in progress to get a more
precise picture of these excitations. In addition, the presented
results will also have dramatic consequences for more realistic models
where the interchange symmetry between spin and orbital degrees of
freedom is only approximately valid. This is left for future
investigation.

We would like to thank G. Felder, B. Normand, F.C. Zhang and
especially T.M. Rice for many fruitful discussions. One of us (B.F.)
is also grateful for financial support from the Swiss Nationalfonds.
The calculations were performed on the Intel Paragon at the ETH
Z\"urich.

\end{document}